\documentclass[aps,prstper,reprint,groupedaddress,showkeys,floatfix,longbibliography]{revtex4-1}
\usepackage{dcolumn}
\usepackage{graphicx}
\usepackage{natbib}
\usepackage{epstopdf}
\usepackage{amsmath}
\usepackage{relsize}
\usepackage{color}
\usepackage{times}
\usepackage[colorlinks=true,allcolors=blue]{hyperref}

\bibpunct{[}{]}{,}{n}{}{}

\renewcommand\arraystretch{1.00}

\begin{document}
\title{Representing uncertainty on model analysis plots}
\author{Trevor I. Smith}
\affiliation{Department of Physics and Astronomy and Department of STEAM Education, Rowan University, Glassboro, New Jersey 08028, USA}


\begin{abstract}
Model analysis provides a mechanism for representing student learning as measured by standard multiple-choice surveys. The model plot contains information regarding both how likely students in a particular class are to choose the correct answer and how likely they are to choose an answer consistent with a well-documented conceptual model. Unfortunately Bao's original presentation of the model plot did not include a way to represent uncertainty in these measurements. I present details of a method to add error bars to model plots by expanding the work of Sommer and Lindell. I also provide a template for generating model plots with error bars. 
\end{abstract}

\maketitle
\section{Introduction}
Model analysis is a powerful tool for representing student learning in terms of both increases in the use of correct models and decreases in the use of incorrect models. Bao and Redish introduced model analysis as a complement to typical representations of learning gains that focus on student correctness \cite{Bao2006}. The model plot simultaneously shows how much a class's use of the correct model increases and how much their use of a well-defined incorrect model decreases (or vice versa). Student use of these models are often measured by a mutiple-choice survey, such as the Force Concept Inventory (FCI) \cite{Hestenes1992} or the Force and Motion Conceptual Evaluation (FMCE) \cite{Thornton1998}. Smith, Wittmann, and Carter have used model analysis in conjunction with statistical analyses of students' normalized gains to compare the effects of various instructional strategies on student learning at several colleges and universities \cite{Smith2014}. Rakkapao, Pengpan, Srikeaw, and Prasitpong also report on the benefits of using model analysis to represent the rich variety of data that come from comparing instructional methods, including cases in which student use of both the correct and common incorrect model increase \cite{Rakkapao2013}.

Smith, Wittmann, and Carter introduced a method for adding error bars to a model plot as a representation for experimental uncertainty \cite{Smith2014}. In this paper I refine this process and provide additional details about the methods and assumptions for generating errors bars. I also provide templates for generating model plots that include error bars using either Mathematica or the R software environment.

\section{Density matrices and the model plot}
The main goal of model analysis is to use response frequencies to determine the probabilities of students in a particular class using each well-defined model. One step is to create a density matrix to represent a class's knowledge state at a given time \cite{Bao2006}. The class density matrix $D$ is the sum of students' individual density matrices, each of which is determined by the measured frequencies of each student using each of the models.
\begin{equation}
\label{sl-data}
D=\mathlarger{\mathlarger{\sum}}_{i=1}^N\left(\begin{array}{cccc}p_{1,i}&\sqrt{p_{1,i}p_{2,i}}&\dots&\sqrt{p_{1,i}p_{n,i}}\\\sqrt{p_{2,i}p_{1,i}}&p_{2,i}&\dots&\sqrt{p_{2,i}p_{n,i}}\\\vdots&&\ddots&\vdots\\\sqrt{p_{n,i}p_{1,i}}&\sqrt{p_{n,i}p_{2,i}}&\dots&p_{n,i}\end{array}\right)
\end{equation}
where $p_{j,i}$ is the probability that the $i$th student uses the $j$th model to answer a particular question. Typically model 1 is the correct Newtonian model (e.g., net force is proportional to acceleration), models 2 through $(n-1)$ are associated with well-documented incorrect models (e.g., net force is proportional to velocity), and model $n$ is the catch-all for any ``other incorrect'' responses.

\begin{figure}[bhtp]
\begin{center}
\includegraphics[height=5cm]{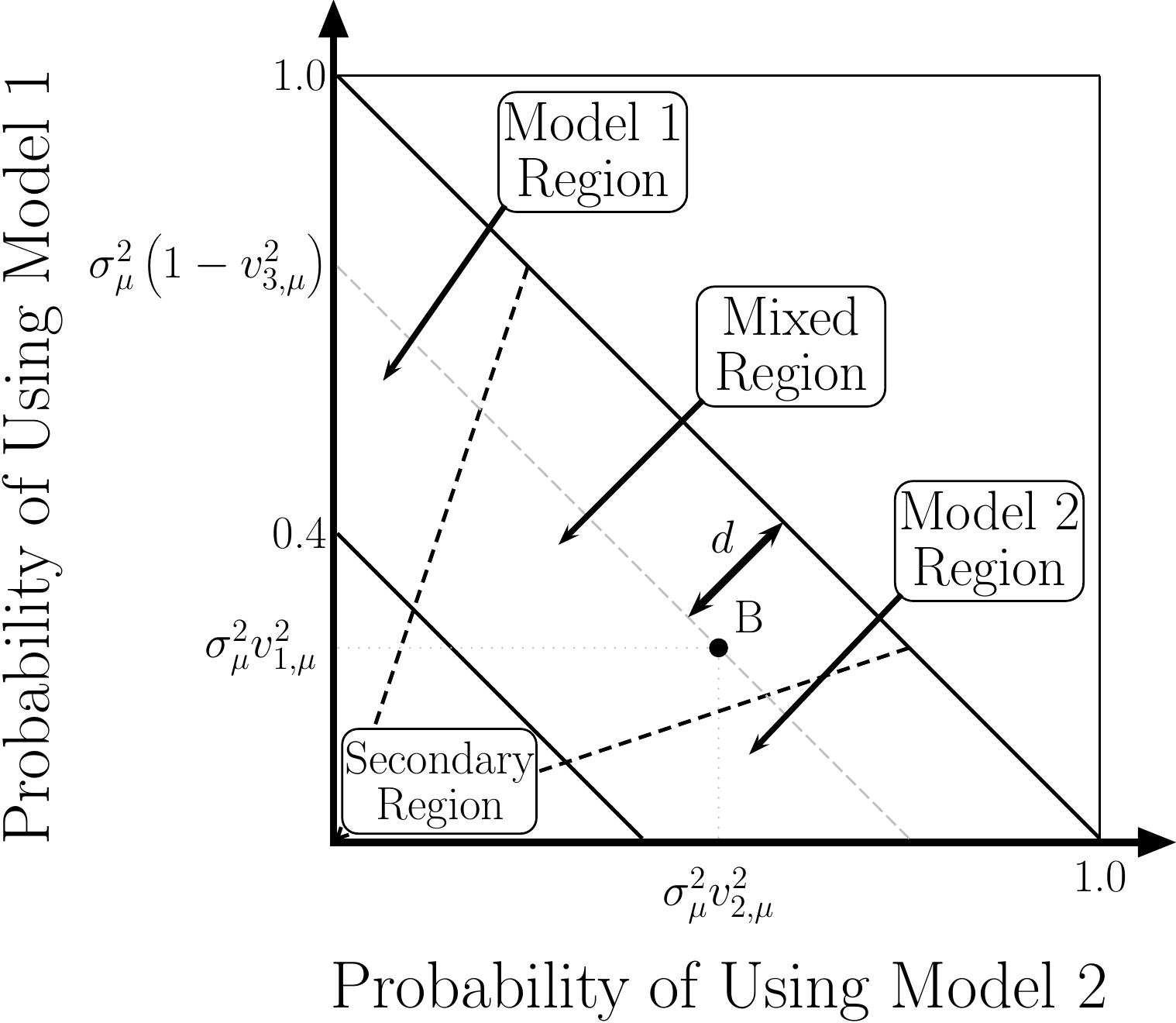}
\caption[Example of a model plot]{Various regions of a model plot with two well-documented mental models. Figure recreated from Ref.\ \cite{Bao2006} and originally published in Ref.\ \cite{Smith2014}, where $\sigma_\mu^2$ is the $\mu$th eigenvalue of the class density matrix, and $v_{k,\mu}$ is the $k$th component of the $\mu$th eigenvector.}
\label{modelplot}
\end{center}
\end{figure}

The eigenvalues and eigenvectors of $D$ are used to characterize the class's knowledge state \cite{Bao2006}. The primary eigenvalue and the components of its associated eigenvector are used to create a single point on a model plot representing a class's probability of using each model at a given time. Figure \ref{modelplot} shows a sample model plot for a physical situation with two well-defined models \cite{Bao2006,Smith2014}.

\section{Uncertainty and the need for error bars}
One shortcoming of model analysis and the model plot is that statistical uncertainty is not represented in the results. Is data point ``B'' in Fig.\ \ref{modelplot} really in the Mixed Region, or could it be in the Model 2 Region? It is impossible to have confidence in the interpretation of a class's ``state'' without error bars on the model plot.

\subsection{Uncertainty in eigenvalues}
Sommer and Lindell recognized the omission of measures of statistical power and proposed a method for determining the uncertainty in the eigenvalues of the class density matrix \cite{Sommer2003}. Their method considers that the measured probability that a student uses a particular model has an associated uncertainty $\epsilon_{j,i}$ that may be positive or negative ($-1\leq\epsilon_{j,i}\leq1$). The real probability may be as high (or low) as $p_{j,i}+\epsilon_{j,i}$. This results in a single-student error matrix, $e_i$,
\begin{equation}
e_i=\left(\begin{array}{cccc}\epsilon_{1,i}&e_{12,i}&\dots&e_{1n,i}\\e_{21,i}&\epsilon_{2,i}&\dots&e_{2n,i}\\\vdots&&\ddots&\vdots\\e_{n1,i}&e_{n2,i}&\dots&\epsilon_{n,i}\end{array}\right)\label{err1}
\end{equation} 
where, 
\begin{equation}
e_{k\ell,i}\equiv\sqrt{\left(p_{k,i}+\epsilon_{k,i}\right)\left(p_{\ell,i}+\epsilon_{\ell,i}\right)}-\sqrt{p_{k,i}p_{\ell,i}}.
\end{equation}

Unfortunately, the uncertainty of a single student choosing a particular model is typically not knowable from data sets of pre- and post-test surveys. Therefore, Sommer and Lindell assume that the error matrix for the class will have the same form as that of Eq.\ \eqref{err1}:
\begin{align}
E&=\left(\begin{array}{cccc}\epsilon_1&E_{12}&\dots&E_{1n}\\E_{21}&\epsilon_2&\dots&E_{2n}\\\vdots&&\ddots&\vdots\\E_{n1}&E_{n2}&\dots&\epsilon_n\end{array}\right)\\[2ex]
E_{k\ell}&\equiv\sqrt{\left(D_{kk}+\epsilon_k\right)\left(D_{\ell\ell}+\epsilon_\ell\right)}-\sqrt{D_{kk}D_{\ell\ell}}.
\end{align}
Where $D_{kk}$ is one of the diagonal elements of the class density matrix.

Given that the error in the measured probability could be positive or negative, each term in $E$ could also be either positive or negative. This information is used to generate a set of specific error matrices. Because the ($n\times n$) density matrix is symmetric, the general error matrix is also symmetric, yielding $2^{n(n+1)/2}$ specific matrices with different combinations of positive and negative terms. By adding each of these specific error matrices to the class density matrix $D$ and computing the eigenvalues of each of the resulting adjusted density matrices, one can determine the upper and lower bounds for each of the eigenvalues \cite{Sommer2003}. We may now be confident that the actual eigenvalue falls within the range $[\sigma_{\mu,min}^2,\sigma_{\mu,max}^2]$.

While this is a step in the right direction, it falls short of providing a mechanism for representing statistical uncertainty within the model plot (the points on which depend on both eigenvalues and the associated eigenvectors). Moreover, this method requires an initial assumption of the values of the uncertainties, $\epsilon_i$, that are used to create the general error matrix. 

\subsection{Creating error bars on the model plot}
To create error bars on the model plot one must translate the uncertainty associated with the eigenvalue of the density matrix to an uncertainty in each dimension of the model plot. As shown in Fig.\ \ref{modelplot} the horizontal coordinate ($x$) corresponds with the probability of choosing Model 2 and is defined as the product of the primary eigenvalue $\sigma_\mu^2$ with the square of the second component of the associated eigenvector $v_{2,\mu}^2$. Similarly, the vertical coordinate ($y$) is associated with Model 1 and the first component of the eigenvector:
\begin{align}
x&=\sigma_\mu^2v_{2,\mu}^2\\
y&=\sigma_\mu^2v_{1,\mu}^2
\end{align}
The uncertainty in each of these coordinates is determined by the uncertainty in the primary eigenvalue and the components of its associated eigenvector, but this relationship is not necessarily straightforward.

Smith, Wittmann, and Carter assumed that the fractional uncertainty in the primary eigenvalue will be the same as the fractional uncertainty of each coordinate \cite{Smith2014},
\begin{equation}
\frac{\Delta_\mu}{\sigma_\mu^2}=\frac{\Delta_x}{x}=\frac{\Delta_y}{y}
\end{equation}
where $\Delta_\mu$, is defined by the upper and lower bounds: 
\begin{equation}
\Delta_\mu=\frac{\left(\sigma_{\mu,max}^2-\sigma_\mu^2\right)+\left(\sigma_\mu^2-\sigma_{\mu,min}^2\right)}{2}.
\end{equation}
The actual coordinates will fall within the ranges $x\pm\Delta_x$ and $y\pm\Delta_y$. However, this assumption causes the error bars to be proportional to the value of the coordinate, which may not accurately reflect the uncertainty in the class's model state, e.g., $\Delta_x=x\left(\Delta_{\mu}/\sigma^2_\mu\right)$.

A less restrictive method is to determine the coordinates $(x_k,y_k)$ for each of the adjusted density matrices by calculating the eigenvalues and eigenvectors of each. The uncertainty represented by the error bars would then be $[x_{min},x_{max}]$ and $[y_{min},y_{max}]$. I provide an example in Sec.\ \ref{sec:ex} that shows the results of each assumption.

\subsection{Choosing an initial estimate of uncertainty}
Sommer and Lindell propose using a single uncertainty for simplicity ($\epsilon=\mathrm{max}\{\epsilon_1,\epsilon_2,\dots,\epsilon_n\}$) but provide no straightforward method for determining an initial estimate \cite{Sommer2003}. There are several options for choosing an initial estimate of the uncertainty based on the pre- and post-test data. The choices I present are based on the standard error of a particular data set. 

One of the simplest choices is to assume that the uncertainty will be the same for all models and will be the same for both pre- and post-test data. To accomplish this I use the pooled standard error in terms of the standard deviations of both the pre- and post-test data sets:
\begin{equation}
\label{se}
\begin{split}
\lefteqn{
\epsilon=\sqrt{\left(\frac{(N_{\textrm{pre}}-1)s_{\textrm{pre}}^2+(N_{\textrm{post}}-1)s_{\textrm{post}}^2}{N_{\textrm{pre}}+N_{\textrm{post}}-2}\right)} }\hspace{4cm} \\
&\times\left(\frac{1}{N_{\textrm{pre}}}+\frac{1}{N_{\textrm{post}}}\right)^{\frac{1}{2}},
\end{split}
\end{equation}
where $N$ is the number of students and $s$ is the standard deviation of the number of correct answers for each data set \footnote{Equation \eqref{se} assumes that student scores are recorded as a percentage correct. If instead scores are recorded directly, then the entire expression should be divided by the number questions in the particular cluster.}.

For cases in which one uncertainty may not fit the data one may choose to calculate the standard error for the pretest and post-test separately:
\begin{align}
\epsilon_{pre}&=\frac{s_{pre}}{\sqrt{N_{pre}}}\label{sepre}\\[1ex] 
\epsilon_{post}&=\frac{s_{post}}{\sqrt{N_{post}}}\label{sepost}.
\end{align}

Additionally, one may choose not to accept the assumption proposed by Sommer and Lindell that the uncertainty is the same for all models. Equations \eqref{se}--\eqref{sepost} may all be applied to the data sets of students using each of the models. In the following section I provide examples for a single value of uncertainty for all models pre- and postinstruction (the most restrictive assumption) and different uncertainties for all models (the least restrictive assumption).

\section{Example of generating error bars}
\label{sec:ex}
\begin{figure*}[tb]
\begin{center}
\includegraphics{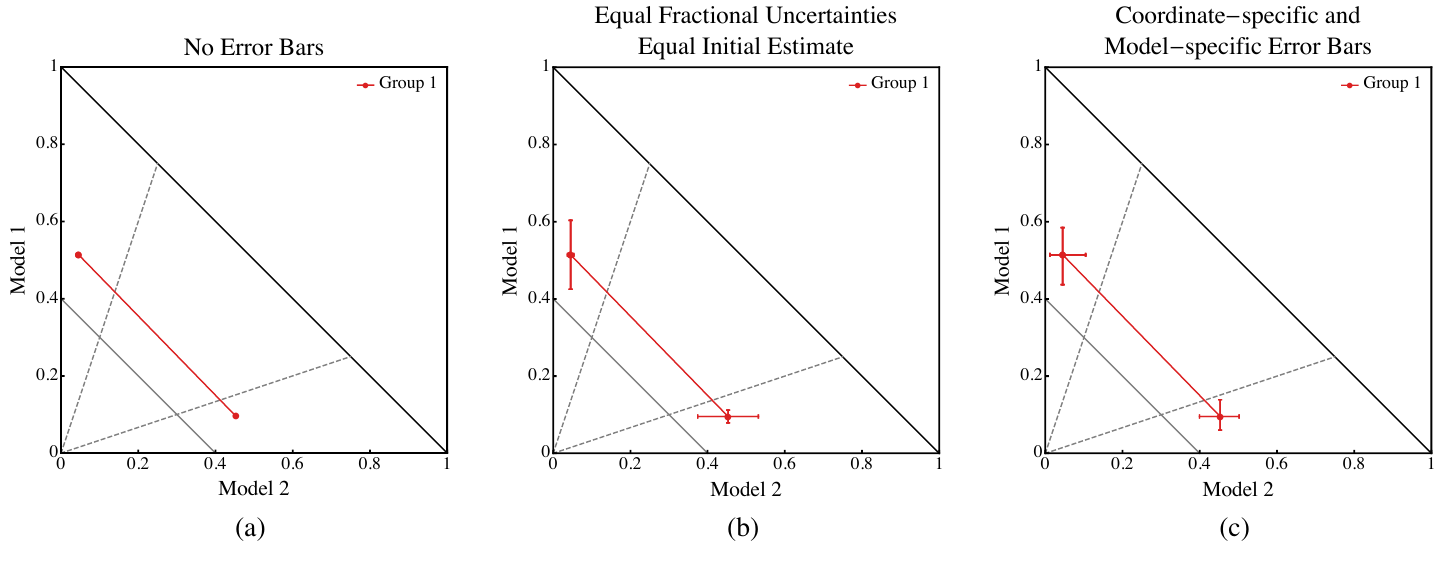}
\caption{Model plots for the example data: (a) no error bars, (b) assuming equal fractional uncertainties for both coordinates and equal initial error estimates for all models (pooled standard error), and (c) assuming coordinate-specific uncertainties and model-specific initial error estimates (standard error of the distribution for each model).}
\label{all-plots}
\end{center}
\end{figure*}

I present an example to illustrate the process of creating error bars on the model plot and to examine the implications of each of the assumptions for choosing an initial error estimate. For the sake of brevity I only present pretest and post-test data for a single class in one year answering questions within a single question cluster \footnote{These data were originally published in Ref.\ \cite{Smith2014}: School 1, Year 2, Energy cluster.}. In this question cluster there are two well-defined models (correct and common incorrect) and one ``other'' model. The pre- and post-test class density matrices are:
\begin{align}
D_{pre}&=\left(\begin{array}{ccc}
0.222&0.162&0.123\\
0.162&0.500&0.270\\
0.123&0.270&0.278\end{array}\right)\\[2ex]
D_{post}&=\left(\begin{array}{ccc}
0.559&0.089&0.140\\
0.089&0.218&0.090\\
0.140&0.090&0.224\end{array}\right)
\end{align}
These density matrices have eigenvalues and associated eigenvectors that yield coordinates on the model plot:
\begin{align}
\sigma_{pre}^2&=0.758&\mathbf{v}_{pre}&=\left(\begin{array}{c}0.355\\0.773\\0.525\end{array}\right)\\[1ex]
x_{pre}&=0.453&y_{pre}&=0.095\\[2ex]
\sigma_{post}^2&=0.641&\mathbf{v}_{post}&=\left(\begin{array}{c}0.896\\0.265\\0.357\end{array}\right)\\[1ex]
x_{post}&=0.045&y_{post}&=0.514.
\end{align}
As seen in Fig.\ \ref{all-plots}(a) this class starts in the Model 2 region and moves to the Model 1 region. In the following sections we examine the implications of each of our assumptions: equal fractional uncertainties vs.\ coordinate-specific uncertainties, and equal initial error estimates vs.\ model-specific errors.

\subsection{Assuming equal fractional uncertainty and equal error estimates}
As a starting point I assume that the fractional uncertainties for both coordinates are the same as that of the primary eigenvalue and that a single estimate of uncertainty will suffice for all models for both the pre- and post-test data \footnote{These are the same assumptions that were used to generate the model plots in Ref.\ \cite{Smith2014}.}. The statistics for each data set are contained in Table \ref{stat}. The pooled standard error for the correct responses is $\epsilon=0.0470$, which is higher than either of the individual standard errors for pre/post data. This provides general error matrices of,
\begin{align}
E_{pre}&=\left(\begin{array}{ccc}
\pm0.0470&\pm0.0502&\pm0.0475\\
\pm0.0502&\pm0.0470&\pm0.0533\\
\pm0.0475&\pm0.0533&\pm0.0470
\end{array}\right)\label{gen-e-1}\\[2ex]
E_{post}&=\left(\begin{array}{ccc}
\pm0.0470&\pm0.0500&\pm0.0570\\
\pm0.0500&\pm0.0470&\pm0.0489\\
\pm0.0570&\pm0.0489&\pm0.0470
\end{array}\right)\label{gen-e-2}
\end{align}
where the ``$\pm$'' indicate that each cell could be positive or negative.

\begin{table}[bt]
\caption{Statistics regarding the number of students ($N$), average values, standard deviations ($s$), and standard errors ($\epsilon$) of Model 1 and Model 2 scores for both the pre- and post-test data.}
\begin{center}
\renewcommand\arraystretch{1.25}
\begin{ruledtabular}
\begin{tabular}{lccccccc}
&&\multicolumn{3}{c}{Model 1: Correct}&\multicolumn{3}{c}{Model 2: Common}\\
&$N$&Average&$s_{M1}$&$\epsilon_{M1}$&Average&$s_{M2}$&$\epsilon_{M2}$\\
\hline
pre&109&0.222&0.264&0.0253&0.500&0.274&0.0263\\
post&85&0.559&0.389&0.0422&0.218&0.311&0.0337\\
pooled&&&&0.0470&&&0.0421\\
\end{tabular}
\end{ruledtabular}
\renewcommand\arraystretch{1.00}
\end{center}
\label{stat}
\end{table}%

Using every combination of positive and negative cells yields 64 adjusted density matrices. For the pretest the eigenvalues span the range $[0.603,0.849]$, and on the post-test they span the range $[0.425,0.614]$. This gives general uncertainties of $\Delta_{\mu,pre}=0.123$ and $\Delta_{\mu,post}=0.094$. Assuming equal fractional uncertainties provides an uncertainty for each coordinate:
\begin{align}
\Delta_{x,pre}&=0.0903&\Delta_{y,pre}&=0.0183\\
\Delta_{x,post}&=0.0108&\Delta_{y,post}&=0.0820
\end{align}
As can be seen in Fig.\ \ref{all-plots}(b) the point in the Model 2 region has a much larger uncertainty in the horizontal coordinate and vice versa. This is a direct result of the assumption of equal fractional uncertainties for which the uncertainty in a coordinate is proportional to the value of that coordinate.

\subsection{Assuming coordinate-specific uncertainty and model-specific error estimates}
We now present the results of rejecting both the assumption of equal fractional uncertainties and the assumption of equal initial error estimates. Using the standard error for each model in each data set (see Table \ref{stat}) provides new general error matrices:
\begin{align}
E_{pre}&=\left(\begin{array}{ccc}
\pm0.0253&\pm0.0276&\pm0.0231\\
\pm0.0276&\pm0.0263&\pm0.0233\\
\pm0.0231&\pm0.0233&\pm0.0202
\end{array}\right)\label{gen-e-3}\\[2ex]
E_{post}&=\left(\begin{array}{ccc}
\pm0.0422&\pm0.0399&\pm0.0368\\
\pm0.0399&\pm0.0337&\pm0.0318\\
\pm0.0368&\pm0.0318&\pm0.0299
\end{array}\right)\label{gen-e-4}
\end{align}
We use these to calculate the $x$ and $y$ coordinates for each of the $2^{n(n+1)/2}$ adjusted density matrices.
\begin{align}
x_{pre}&\in[0.399,0.502]\\
y_{pre}&\in[0.060,0.139]\\
x_{post}&\in[0.012,0.104]\\
y_{post}&\in[0.437,0.585]
\end{align} 

As shown in Fig.\ \ref{all-plots}(c) this gives a much different representation of the uncertainty in the coordinates on the model plot. This provides a more accurate representation of the span of the model space each point could occupy given that these uncertainties do not depend directly on the values of the coordinates themselves. These error bars also show that the uncertainty in the post-test data is greater than in the pretest data (see Table \ref{stat}).

\section{Summary}
The assumptions involved in calculating error bars can have a dramatic effect on the interpretations reflected in the model plot. The sample data above show that using the standard error of the data yields error bars with a non-negligible extent on the model plot. This is most visible in Fig.\ \ref{all-plots}(c) where the model-specific error estimates and coordinate-specific uncertainties provide macroscopic error bars in both coordinates for both points. These results support the notion that it is imperative to represent the uncertainty in the coordinates on the model plot in some fashion. The class model state is not precisely known as would be implied otherwise.

In an effort to facilitate the use of this method I have included several template files in the supplemental materials that may be used to generate density matrices and model plots with error bars \footnote{See Supplemental Material at [URL will be inserted by publisher].}. The Excel file (\verb|MA_FMCE_template.xltx|) generates class density matrices from student responses to the FMCE \footnote{The Excel template also works in OpenOffice.}. This file may be modified in order to create density matrices for any other multiple choice data with well-defined models. The text files include the commands for importing data from Excel, performing the necessary matrix calculations, and generating model plots with error bars using the open-source R software environment \footnote{Excel files must have the \texttt{.xlsx} extension to be imported into R.}. The file \verb|MA_3Model_templateR.txt| assumes a question cluster with three models (two well-defined models and one ``other incorrect'' as is the case with the above example), and the file \verb|MA_4Model_templateR.txt| assumes a question cluster with four models \footnote{The two additional files \texttt{MA\_3Model\_templateR-ods.txt} and \texttt{MA\_4Model\_templateR-ods.txt} are designed to import data from OpenOffice (\texttt{.ods}) files rather than Excel files.}. I have also included Mathematica template files (\verb|MA_3Model_template.nb| and \verb|MA_4Model_template.nb|) that perform the same functions as the R files but are not as flexible.  All files include instructions for creating model plots with error bars starting from a class set of multiple-choice survey data.

\begin{acknowledgments}
I thank Michael C.\ Wittmann for productive conversations about model analysis, and Ian Griffin, Nick Wright, and Ashley Smith for testing the analysis templates. I am also grateful to an anonymous reviewer on Ref.\ \cite{Smith2014} for bringing my attention to the work of Sommer and Lindell. This research was partially supported by the Rowan University Seed Funding Program.
\end{acknowledgments}

\bibliography{TIS}

\end{document}